\documentclass[twocolumn,showpacs,aps,pre,amsmath,amssymb,tightenlines,10pt]{revtex4}
\usepackage{graphicx}
\usepackage{dcolumn}
\usepackage{bm}

\def\be{\begin{equation}}
\def\ee{\end{equation}}
\def\bea{\begin{eqnarray}}
\def\eea{\end{eqnarray}}
\def\ba{\begin{array}}
\def\ea{\end{array}}

\def\0{$\Gamma_0$}

\begin{document}

\title{Soluble kagome Ising model in a magnetic field}
\author{W. T. Lu$^{1,2}$ and F. Y. Wu$^{1}$}
  \affiliation{$^1$Department of Physics, $^2$Electronic Materials Research Institute,
  Northeastern University, Boston, Massachusetts 02115}
\date{\today}

\begin{abstract}
An Ising model on the kagome lattice with super-exchange interactions is
solved exactly under the presence of a nonzero external magnetic field. The
model generalizes the super-exchange model introduced by Fisher in 1960 and
is analyzed in light of a free-fermion model. We deduce the critical
condition and present detailed analyses of its thermodynamic and magnetic
properties. The system is found to exhibit a second-order transition with
logarithmic singularities at criticality.
\end{abstract}


\maketitle

\section{Introduction}

The Ising model in a nonzero magnetic field is a well-known unsolved problem
in statistical mechanics. In 1960, Fisher \cite{Fisher60} produced a
remarkable solution of a super-exchange antiferromagnetic Ising model in the
presence of a nonzero field. The Fisher model is defined on a decorated
square lattice where there is an external magnetic field applied to
decorating spins which interact via super-exchange interactions.

\begin{figure*}[htbp]
\vskip 5mm
\center{
\includegraphics [angle=0,width=12cm]{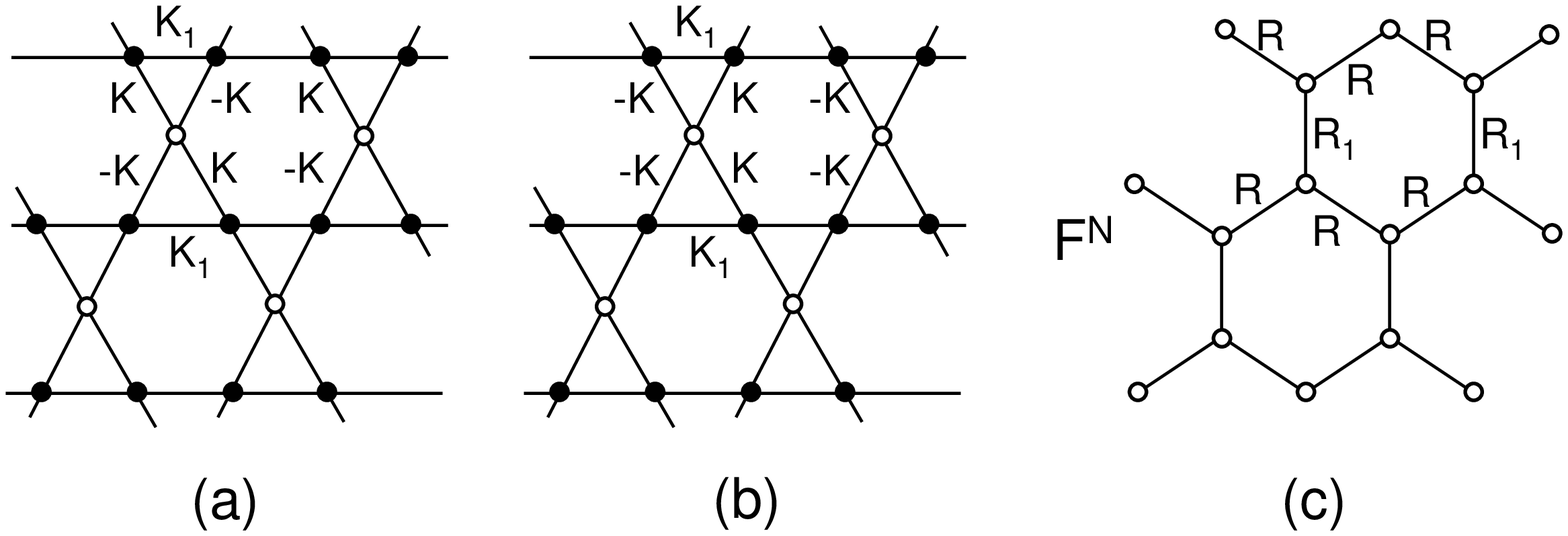}}
\caption{(a) and (b) are two soluble Kagome lattice Ising models in a magnetic field. 
Solid circles denote magnetic spins
and open circles non-magnetic spins. Both models are mapped to an honeycomb
Ising lattice without a field shown in (c). $F^N$ is an overall factor.}
\label{fig0}
\end{figure*}

In this paper we consider similar super-exchange models on the kagome
lattice. The kagome lattice has been of interest in recent years in the role
it plays in the high-$T_{c}$ superconductivity. 
It has been known that special cases  of the Kagome Ising model 
are soluble in the presence of a magnetic field \cite{Giacomini88,Lin89}.
The structure of the
ferrimagnet SrCr$_{8}$Ga$_{4}$O$_{19}$, for example, is found to consist of
2D spinel (kagome) slabs \cite{Ob} with magnetic spins residing at 1/3 of
the lattice sites. Thus it is  of interest
to consider models with similar structures.
   Azaria and Giacomini \cite{Azaria88} have extended
the Fisher model by considering the kagome lattice shown in Fig. \ref{fig0}(a),
which reduces to the Fisher model upon setting $K_1=0$.  They obtained its
partition function from which the phase diagram is deduced. But there has been
no detailed discussion of its thermodynamic properties.
Here, we consider yet another extension of the Fisher model to a kagome
lattice  as shown in Fig. \ref{fig0}(b).  We show that the partition function
of this model is identical to the Azaria and Giacomini solution \cite{Azaria88},
although there appears to be no direct mapping between the two models.
In addition, we also present detailed analyses of the thermodynamics of the solution.

Consider an Ising model 
shown in Fig. \ref{fig0}(b) with an interaction energy 
\begin{eqnarray}
-J_{1}\sigma _{1}\sigma _{2}+J(\sigma _{2}\sigma _{3}-\sigma _{3}\sigma _{1})\nonumber
\end{eqnarray}%
around every triangle formed by spins $\sigma _{1},\sigma _{2},\sigma _{3}$.
Introduce reduced interactions $K=\beta J,K_{1}=\beta J_{1}$, where $\beta
=1/kT$, such that $K,K_{1}>0$ indicate ferromagnetic interactions. In
addition, there is an external magnetic field $H$ applied to 2/3 of the
lattice sites denoted by solid circles. We denote the reduced field by $%
L=H/kT$. As a result, the magnetic spins interact with an super-exchange
interaction via intermediate non-magnetic spins. It is clear that there is
no loss of generality to restrict considerations to 
\begin{eqnarray}
H,J\geq 0,\quad \mathrm{or}\quad L,K\geq 0.\nonumber
\end{eqnarray}

For $K_1 = 0$ the model reduces to the Fisher model for the square lattice.
For $K_1 \not= 0$ the present model is more general and differs from the
Fisher model in a fundamental way as described in Sec. 2 below.

Our main result is a closed-form solution of the partition function and 
detailed analyses of thermodynamic and magnetic properties. In Sec.
2 we deduce the solution using a combination of star-triangle and decimation
transformations. The phase diagram is analyzed in Sec. 3. In Secs. 4 - 6 the
internal energy, specific heat, magnetization, and susceptibility are
analyzed using a free-fermion model formulation \cite{Fan70}.

\section{The partition function}

Denote the partition function of the kagome Ising lattice in Fig. 1(b) 
by $Z_{\mathrm{KG}}(K,K_1,L)$. Our main result is an equivalence of 
$Z_{\mathrm{KG}}(K,K_1,L)$ to the partition function of an honeycomb 
Ising model in zero field, a result which renders the model soluble.

This equivalence is established by effecting a sequence of spin
transformations. First, we carry out a star-triangle transformation for
every triangular face as shown in the first line in Fig. 2. This converts
the kagome lattice to a decorated honeycomb lattice. Next the decorating
spins are decimated as shown in the second and third lines in Fig. 2, and
the lattice is reduced to that of an honeycomb. The crux of matter is that
the external fields $L_1$ and $-L_1$ induced in the second step cancel out
at the end. As a result, the final honeycomb lattice has \textit{no}
external field and is soluble. It has reduced interactions $R,R$ and $R_1$
in the three principal directions.

\begin{figure}[htbp]
\vskip 5mm
\center{
\includegraphics [angle=0,width=6cm]{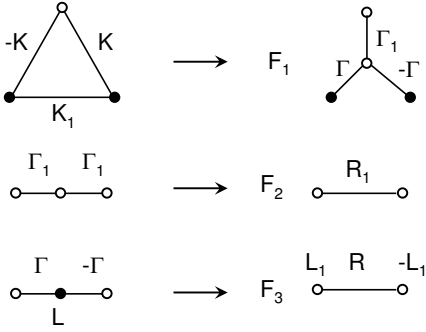}}
\caption{Transformations used in
deducing the mapping shown in Fig. 1. $F_{i},i=1,2,3$ are constants.}
\label{fig2}
\end{figure}

The transformations shown in Fig. 2 are standard \cite{Syozi,Syozi1}. For
the transformation in the first line in Fig. 2, we have 
\begin{eqnarray}
e^{-2K-K_1} &=& 2 F_1 \cosh( \Gamma_1+2\Gamma)  \label{11} \\
e^{2K-K_1} &=& 2 F_1 \cosh( \Gamma_1-2\Gamma)  \label{12} \\
e^{K_1} &=& 2 F_1 \cosh \Gamma_1  \label{13}
\end{eqnarray}
from which we can solve $\Gamma$ and $\Gamma_1$ in terms of $K$ and $K_1$.
Divide the sum and difference of (\ref{11}) and (\ref{12}) by (\ref{13}), we
obtain, respectively, 
\begin{eqnarray}
\cosh 2\Gamma &=& e^{-2K_1}\cosh 2K  \label{14} \\
\tanh \Gamma_1 &=& - \frac {e^{-2K_1}\sinh 2K}{\sinh 2\Gamma}.  \label{15}
\end{eqnarray}
We also obtain 
\begin{equation}
F_1^4 =\big[16 \cosh^2\Gamma_1 {\cosh(\Gamma_1+2\Gamma)} {%
\cosh(\Gamma_1-2\Gamma)} \big]^{-1}.  \label{f1}
\end{equation}
Now $K>0$ and $K_1$ real, so (\ref{14}) and (\ref{15}) show that there are
two regimes 
\begin{eqnarray}
e^{-2K_1}\cosh 2K >1&& \quad \mathrm{regime\>\>I}  \notag \\
e^{-2K_1}\cosh 2K <1 && \quad \mathrm{regime\>\>II}  \label{regime}
\end{eqnarray}
and $\Gamma$ and $\Gamma_1$ are real in regime I and pure imaginary in
regime II.

For the transformation in the second line in Fig. 2, we have 
\begin{eqnarray}
2 \cosh 2\Gamma_1 &=& F_2\, e^{R_1}  \notag \\
2 &=& F_2 \, e^{-R_1}  \label{22}
\end{eqnarray}
from which we obtain 
\begin{eqnarray}
F_2^2 &=& 4\cosh 2\Gamma_1  \label{f2} \\
e^{2R_1} &=& \cosh 2\Gamma_1 = \frac {e^{-4K_1} \cosh 4K -1}{e^{-4K_1} -1}
\label{23}
\end{eqnarray}
where the last step is obtained after making use of (\ref{14}). Equation (%
\ref{23}) shows that $R_1$ is real, and $R_1>0$ in regime I and $R_1<0$ in
regime II.

For the transformation in the third line in Fig. 2, we have 
\begin{eqnarray}
2\cosh (L+2\Gamma ) &=&F_{3}\,e^{-R+2L_{1}}  \label{31} \\
2\cosh (L-2\Gamma ) &=&F_{3}\,e^{-R-2L_{1}}  \label{32} \\
2\cosh L &=&F_{3}\,e^{R}  \label{33}
\end{eqnarray}%
from which we obtain 
\begin{eqnarray}
e^{-4R} &=&\frac{\cosh (L+2\Gamma )\cosh (L-2\Gamma )}{\cosh ^{2}L}=1+\frac{%
\sinh ^{2}2\Gamma }{\cosh ^{2}L}  \notag \\
&=&\frac{e^{-4K_{1}}\cosh ^{2}2K+\sinh ^{2}L}{\cosh ^{2}L}  \label{34} \\
e^{4L_{1}} &=&\frac{\cosh (L+2\Gamma )}{\cosh (L-2\Gamma )}  \label{35} \\
F_{3}^{4} &=&16\cosh ^{2}L{\cosh (L+2\Gamma )}{\cosh (L-2\Gamma )}.
\label{f3}
\end{eqnarray}%
Thus, $R$ is always real, and we have 
\begin{eqnarray}
R<0,R_{1}>0, &&\Gamma ,\Gamma _{1},L_{1}=\mathrm{real},\hskip1cm\mathrm{%
regime\>\>I}  \notag \\
R>0,R_{1}<0, &&\Gamma ,\Gamma _{1},L_{1}=\mathrm{imaginery},\mathrm{%
regime\>\>II}.
\end{eqnarray}%
This says that $R$ and $R_{1}$ have opposite signs. In the solution obtained by 
Azaria and Giacomini \cite{Azaria88}, the final equivalent honeycomb Ising model 
is identical with ours except the replacement of $R$ by $-R$. However, as we 
shall seen (\ref{fkg}) below, the negations of $R$ and/or $R_{1}$ do not 
affect the solution. Therefore, the two solutions are identical.


Combining (\ref{f1}), (\ref{f2}), (\ref{f3}), and after some reduction, we
obtain 
\begin{eqnarray}
F^{4} &=&F_{1}^{4}F_{2}^{4}F_{3}^{2}  \notag \\
&=&\frac{4\cosh ^{2}L(e^{4K_{1}}\sinh ^{2}L+\cosh ^{2}2K)}{\cosh ^{2}R_{1}}.
\label{R-L1}
\end{eqnarray}%
Now the field $L_{1}$ at the decorating sites is canceled and we arrive at
the final equivalence 
\begin{equation}
Z_{\mathrm{KG}}(K,K_{1},L)=F^{N}\ Z_{\mathrm{HC}}[R(L),R_{1}]  \label{equiv}
\end{equation}%
where $Z_{\mathrm{HC}}[R(L),R_{1}]$ is the partition function of the
honeycomb Ising lattice. Here, $N$ is the number of triangular faces (or the
number of magnetic spins) of the kagome lattice, which is also the number of
honeycomb lattice sites. Note that the $L$ dependence is only in $R$.

The transformation (\ref{equiv}) reduces to that of Fisher's for the square
lattice after setting $K_1=0\ (R_1=\infty)$. But our model differs from the
Fisher model in a crucial aspect: The Fisher model results in a square
lattice whose sites are the decorating sites of the original lattice, in the
present model all sites of the original kagome lattice disappear at the end.
Thus the ordering of the honeycomb lattice bears no direct relationship to
that of the kagome lattice. In addition, while there is no frustrated
plaquettes in the Fisher model, in the $K_1>0$ model all triangular faces
are frustrated.

Introduce the \textit{per magnetic spin} free energy 
\begin{eqnarray}
\hskip 5mm f_{\mathrm{KG}}(K,K_1,L)&=&\lim_{N\to \infty} N^{-1}\ln Z_{%
\mathrm{KG}}(K,K_1,L)  \notag
\end{eqnarray}
and the per-site honeycomb lattice free energy 
\begin{eqnarray}
f_{\mathrm{HC}}(R,R_1)&=&\lim_{N\to \infty} {\ N}^{-1}\ln Z_{\mathrm{HC}%
}(R,R_1).  \label{hcfree}
\end{eqnarray}
Taking the $N\to \infty$ limit of (\ref{equiv}) and making use of the
explicit expression of $f_{\mathrm{HC}}$ given in \cite{Syozi1,Houtappel},
we obtain for the kagome Ising model, 
\begin{eqnarray}
f_{\mathrm{KG}}(K,K_1,L) &=& \ln F + f_{\mathrm{HC}}(R,R_1)  \notag \\
&=& \ln F +{\frac{3}{4}}\ln 2  \notag \\
&&\quad +{\frac{1}{16 \pi^2}}\int_0^{2\pi}\!\!\!d\theta d\phi\ln \Xi
(\theta,\phi)  \label{freeenergy}
\end{eqnarray}
where 
\begin{eqnarray}
\Xi (\theta,\phi) &=& \cosh 2R_1 \cosh^22R + 1-\sinh^22R\cos(\theta+\phi) 
\notag \\
&&-\sinh2R_1\sinh2R(\cos\theta+\cos\phi).  \label{xi}
\end{eqnarray}
Note that the negation of either $R$ or $R_1$ corresponds to changing $%
\theta\to \pi-\theta$, $\phi \to \pi-\phi$ in (\ref{xi}) which does not
change the free energy $f_{\mathrm{KG}}$.

As a check, we recover the Fisher solution \cite{Fisher60} upon setting $%
K_{1}=0$ or $R_{1}\rightarrow \infty $: 
\begin{eqnarray}
&&\hskip-5mmf_{\mathrm{KG}}(K,0,L)={\frac{3}{2}}\ln 2+{\frac{1}{4}}\ln
[\cosh ^{2}L(\sinh ^{2}L+\cosh ^{2}2K)]  \notag \\
&&\hskip6mm+{\frac{1}{16\pi ^{2}}}\int_{0}^{2\pi }\!\!\!d\theta d\phi \ln %
\Big[\cosh ^{2}2R  \notag \\
&&\hskip38mm-\sinh 2R(\cos \theta +\cos \phi )\Big]  \label{fkg}
\end{eqnarray}%
with $e^{-4R}=(\sinh ^{2}L+\cosh ^{2}2K)/\cosh ^{2}L$.

The internal energy and magnetization per magnetic spin are, respectively, 
\begin{eqnarray}
U(K, K_1, L) &=& -\frac \partial {\partial \beta} f_{\mathrm{KG}}(K,K_1,L),
\label{energy} \\
M(K, K_1, L) &=& \frac \partial {\partial L} f_{\mathrm{KG}}(K,K_1,L).
\label{mag}
\end{eqnarray}
The specific heat and susceptibility are further computed as 
\begin{eqnarray}
C_H (K, K_1, L) &=& \frac \partial {\partial T} U(K, K_1, L) ,
\label{specificheat} \\
\chi (K, K_1, L) &=& \frac \partial {\partial H} M(K, K_1, L).
\label{susceptibility}
\end{eqnarray}

\section{The phase diagram}

The integral in the free energy $f_{\mathrm{KG}}$ is precisely of the form
of that of the free-fermion model discussed by Fan and Wu \cite{Fan70} for
which the critical condition is (for $RR_1<0$) 
\begin{equation}
\Xi(\pi,\pi)=0.  \label{hccritical}
\end{equation}
Alternately, the critical point for an anisotropic honeycomb Ising lattice
has been given in \cite{Houtappel} as $C_1C_2C_3+1 = S_1S_2+S_2S_3+S_3S_1$
where $C_i = \cosh 2R_i, S_i = \sinh |2R_i|, i=1,2,3$. It is also given by
the expression $\omega_1=\omega_2+\omega_3+\omega_4$ \cite{Fan70} where the $%
\omega$'s are given in the next section. Using any of these expressions, we
obtain the critical condition 
\begin{equation}
\sinh [2R(L)] =- \coth R_1.  \label{hccritical1}
\end{equation}
Explicitly, (\ref{hccritical1}) reads 
\begin{eqnarray}
\cosh^2L&=&\frac{1}{2} \Big[\sqrt{1+\tanh^2 R_1}-1\Big]  \notag \\
&&\times \Big( {\ e^{-4K_1}\cosh^22K-1}\Big), \quad T=T_c(H)
\label{hccritical2}
\end{eqnarray}
which can be realized only in regime I and $K_1 \leq K$, or $\gamma \leq 1$
(see below). Note that for $K_1=0$ (\ref{hccritical2}) reduces to the Fisher
expression 
\begin{equation}
\cosh L = \sqrt{(\sqrt{2}-1)/ 2}\ \sinh2K\ .
\end{equation}
It can be verified that we have 
\begin{equation}
|\sinh 2R(L)| < |\coth R_1|, \hskip 1cm T>T_c(H).
\end{equation}

Introduce parameters 
\begin{eqnarray}
\alpha &=& L/2K=H/2J >0  \notag \\
\gamma&=&J_1/J = K_1/K  \label{gamma}
\end{eqnarray}
such that $\gamma>0 $ indicates $J_1 $ is ferromagnetic, and consider the
phase diagram (\ref{hccritical2}) in the $\{\alpha, 1/K\}$ plane, where $1/K$
is the temperature. The phase diagram is plotted in Fig. 3 for different
fixed values of $\gamma$.

\begin{figure}[htbp]
\center{
\includegraphics [angle=0,width=8cm]{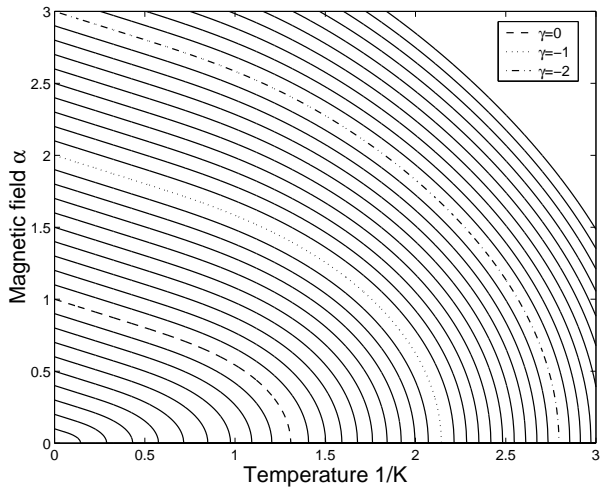}}
\caption{ Phase boundary in the $\protect\alpha $-$T$ plane for fixed values
of $\protect\gamma $.}
\label{fig3}
\end{figure}

At low temperatures the phase boundary behaves as 
\begin{equation}
\alpha =1-\gamma -\frac{1}{4K}\ln 2(\sqrt{2}+1)  \label{K_c-big}
\end{equation}%
with an initial slope independent of $\gamma $. Solving (\ref{hccritical2})
for $\gamma $ at $H=0$, we obtain 
\begin{equation}
\gamma ={\frac{1}{2K_{c}}}\Big[2\ln (\sinh 2K_{c})-\ln (2\cosh 2K_{c})\Big].
\end{equation}%
We see that $T_{c}(0)$ decreases with $\gamma $, reaching $T_{c}(0)=0$ at $%
\gamma =1$. It follows that there is no transition when 
\begin{equation}
\gamma \geq 1,\quad \mathrm{or\>\>}K_{1}\geq K
\end{equation}%
as aforementioned. This can be understood physically since, in the limit of $%
\gamma \rightarrow \infty $, the non-magnetic spins can assume values $\pm 1$
randomly and there is no ordered state.

For small $H$ (\ref{hccritical2}) is expanded as 
\begin{equation}
T_c(H) = T_c(0) - CH^2,  \label{TcH}
\end{equation}
where $C$ is a constant given by 
\begin{eqnarray}
C^{-1}=kJ\Bigg[\frac{32(16-24t^4+t^8)}{t^9(4+4t^2-t^4)} +\frac{
16(-4\gamma+4t+t^5)}{16-t^{8}}\Bigg]  \notag
\end{eqnarray}
with $t=\tanh 2 J/kT_c(0)$.

To determine the nature of regimes in the phase diagram, we consider the
ground state. At zero temperature the kagome lattice can assume two ordered
states:

(i) All magnetic spins aligned in the same direction with the total energy 
\begin{equation}
U_1 = -N(2\alpha + \gamma )J
\end{equation}
independent of the spin directions of the non-magnetic spins. Since the $N/2$
non-magnetic spins can point in any direction, the entropy of this ground
state is $\frac 1 2 Nk\ln 2$.

(ii) The magnetic spins connected by interactions $-J$ assume the value $+1$
(or $-1$) and those connected by interactions $+J$ assume the value $-1$ (or 
$+1$), while all non-magnetic spins are fixed at $+1$ (or $-1$). This forms
a super-exchange antiferromagnetic state with the energy 
\begin{equation}
U_2 = -N(2 -\gamma )J,
\end{equation}
and the ground state is two-fold degenerate with zero entropy in the
thermodynamic limit.

Comparing $U_{1}$ with $U_{2}$, for $\alpha >1-\gamma $ which is the regime
to the \textquotedblleft exterior\textquotedblright\ of the critical curve (%
\ref{hccritical2}), we have $U_{1}<U_{2}$ indicating the system is
paramagnetic. For $\alpha <1-\gamma $ which is the regime enclosed by the
critical curve, we have $U_{1}>U_{2}$ and the system assumes a
super-exchange antiferromagnetic ordering.

\section{Thermodynamic properties}

To analyze thermodynamic and magnetic properties given by (\ref{energy}) - (%
\ref{susceptibility}), we make use of results of \cite{Fan70}. Writing (\ref%
{xi}) in the form of Eq. (16) of \cite{Fan70}, we have 
\begin{equation}
\Xi (\theta ,\phi )=2a+2b\cos \theta +2c\cos \phi +2d\cos (\theta +\phi ),
\end{equation}%
with 
\begin{eqnarray}
a &=&\cosh 2R_{1}\cosh ^{2}2R+1={\frac{1}{2}}(\omega _{1}^{2}+\omega
_{2}^{2}+\omega _{3}^{2}+\omega _{4}^{2}),  \notag \\
b &=&-\sinh 2R_{1}\sinh 2R=\omega _{1}\omega _{3}-\omega _{2}\omega _{4}, 
\notag \\
c &=&-\sinh 2R_{1}\sinh 2R=\omega _{1}\omega _{4}-\omega _{12}\omega _{3}, 
\notag \\
d &=&-\sinh ^{2}2R=\omega _{3}\omega _{4}-\omega _{1}\omega _{2},
\end{eqnarray}%
or 
\begin{eqnarray}
\omega _{1} &=&\cosh (2R-R_{1})  \notag \\
\omega _{2} &=&\cosh (2R+R_{1})  \notag \\
\omega _{3} &=&\omega _{4}\ =\cosh R_{1}.  \label{vertexw}
\end{eqnarray}%
Using (\ref{vertexw}) we compute the parameters $\{x,y,z\}$ introduced in 
\cite{Fan70} as 
\begin{eqnarray}
x &=&\omega _{1}\omega _{4}-\omega _{2}\omega _{3}=-\sinh 2R\sinh 2R_{1} 
\notag \\
y &=&\frac{1}{2}\big(\omega _{1}^{2}-\omega _{2}^{2}-\omega _{3}^{2}+\omega
_{4}^{2}\big)=-\frac{1}{2}\sinh 4R\sinh 2R_{1}  \notag \\
z &=&\omega _{1}\omega _{4}+\omega _{2}\omega _{3}=2\cosh 2R\cosh ^{2}R_{1}.
\label{xyz}
\end{eqnarray}%
Here we have reversed the sign of $x$ from \cite{Fan70} to make $x>0$. This is permitted since
only $x^2$ appears in the ensuing discussions. The critical condition (\ref{hccritical1})
is equivalent to $y=z$ and we have the regimes 
\begin{eqnarray}
&&y>z>x>0,\quad \quad T<T_{c}(H)  \notag \\
&&z>y>x>0,\quad \quad T>T_{c}(H).  \label{regimes}
\end{eqnarray}

It was established in \cite{Fan70} that derivatives of the free energy are
best computed by first carrying out one-fold integration. Adopting notations
in \cite{Fan70}, we obtain after differentiating (\ref{freeenergy}) the
expression 
\begin{equation}
\lbrack f_{\mathrm{KG}}(K,K_{1},L)]^{\prime
}=C_{0}+C_{1}I_{1}+C_{2}I_{2}+C_{3}I_{3},  \label{derivative}
\end{equation}%
where the prime denotes the derivative with respective to some variable such
as $T$ and $H$, 
\begin{eqnarray}
C_{0} &=&\frac{F^{\prime }}{F}+{\frac{b^{\prime }}{4b}},\>\quad
\>|R_{1}|>|R|\quad (b^{2}>d^{2}),  \notag \\
&=&\frac{F^{\prime }}{F}+{\frac{d^{\prime }}{4d}},\>\quad \>|R_{1}|<|R|\quad
(d^{2}>b^{2}),  \notag \\
C_{1} &=&a^{\prime }-{\frac{a}{2}}\bigg({\frac{b^{\prime }}{b}}+{\frac{%
d^{\prime }}{d}}\bigg)-{\frac{b^{2}-d^{2}}{4d}}\bigg({\frac{b^{\prime }}{b}}-%
{\frac{d^{\prime }}{d}}\bigg),  \notag \\
C_{2} &=&{\frac{b}{2}}\bigg({\frac{b^{\prime }}{b}}-{\frac{d^{\prime }}{d}}%
\bigg),  \notag \\
C_{3} &=&-{\frac{1}{2}}\bigg({\frac{b^{\prime }}{b}}-{\frac{d^{\prime }}{d}}%
\bigg){\frac{b^{2}-d^{2}}{2d}}\bigg({\frac{2ad}{b^{2}+d^{2}}}-1\bigg),
\label{coeff-C}
\end{eqnarray}%
and 
\begin{eqnarray}
I_{1} &=&{\frac{1}{8\pi }}\int_{0}^{2\pi }d\phi \lbrack Q(\phi )]^{-1/2}, 
\notag \\
I_{2} &=&{\frac{1}{8\pi }}\int_{0}^{2\pi }d\phi \cos \phi \lbrack Q(\phi
)]^{-1/2},  \notag \\
I_{3} &=&{\frac{1}{8\pi }}\int_{0}^{2\pi }d\phi (1+\omega \cos \phi
)^{-1}[Q(\phi )]^{-1/2}  \label{I123}
\end{eqnarray}%
with 
\begin{eqnarray}
\omega &=&-2bd/(b^{2}+d^{2})>0  \notag \\
Q(\phi ) &=&x^{2}(\cos \phi -yz/x^{2})^{2}  \notag \\
&&+(x^{2}-y^{2})(z^{2}-x^{2})/x^{2}.
\end{eqnarray}

Carrying out the integrations in (\ref{I123}), we obtain explicitly for $%
T\geq T_{c}(H)$ ($z>y>x>0$) the result \cite{corrections} 
\begin{eqnarray}
I_{1} &=&{\frac{1}{2\pi (z^{2}-x^{2})^{1/2}}}K(k),  \notag \\
I_{2} &=&{\frac{1}{2\pi yz(z^{2}-x^{2})^{1/2}}}\Big[z^{2}K(k)+(y^{2}-z^{2})%
\Pi (r,k)\Big],  \notag \\
I_{3} &=&{\frac{1}{2\pi (y+\omega z)(z^{2}-x^{2})^{1/2}}}\Big[yK(k)  \notag
\\
&&\hskip3cm+{\frac{\omega (z^{2}-y^{2})}{z+\omega y}}\Pi (s,k)\Big]
\label{int_I1}
\end{eqnarray}%
where 
\begin{eqnarray}
k^{2} &=&(y^{2}-x^{2})/(z^{2}-x^{2}),  \notag \\
r &=&y^{2}/z^{2},  \notag \\
s &=&(\omega z+y)^{2}/(z+\omega y)^{2}  \label{krs}
\end{eqnarray}%
and $K$ and $\Pi $ are elliptical integrals of the first and third kinds,
respectively, 
\begin{eqnarray}
K(k) &=&\int_{0}^{\pi /2}{\frac{{d\alpha }}{{\ \sqrt{1-k^{2}\sin ^{2}\alpha }%
}}}  \notag \\
\Pi (r,k) &=&\int_{0}^{\pi /2}{\frac{d\alpha }{(1-r\sin ^{2}\alpha )\sqrt{%
1-k^{2}\sin ^{2}\alpha }}}.
\end{eqnarray}%
For $T\leq T_{c}(H)$ ($y>z>x>0$) we obtain the same result with $y$ and $z$
interchanged.

The apparent non-analyticity of $C_{0}$ at $|R_{1}|=|R|$ as indicated in (%
\ref{coeff-C}) is spurious. It can be shown that the combination of $%
C_{0}+C_{3}I_{3}$ is always analytic.

At the transition temperature $y=z$ (or $k=1,r=1$) both $K$ and $\Pi $
diverge as $\ln |T-T_{c}(H)|$. Applying (\ref{derivative}) to the internal
energy (\ref{energy}) where the derivatives are with respect to $T$, we
obtain 
\begin{eqnarray}
&&\hskip-5mm{U}(H)={U}_{c}(H)+\mathrm{const}\>|T-T_{c}(H)|\ln |T-T_{c}(H)|, 
\notag \\
&&\hskip5cmT\rightarrow T_{c}(H).  \label{genU}
\end{eqnarray}%
A further derivative of $U(H)$ as given by (\ref{specificheat}) gives the
specific heat $C_{H}$. Thus, the energy is continuous at $T_{c}(H)$ while
the specific heat diverges logarithmically. These findings, which are the
same as those found in the Fisher model, indicate the occurrence of a second
order transition along the phase boundary in Fig. 3. We plot $U(H)$ and $%
C_{H}$ respectively in Figs. 4 and 5. For completeness, we derive the
explicit expression for $U_{c}(H)$ in the next section.

In the case of zero magnetic field $H=\alpha =0$ and $\gamma =-1$, the
internal energy assumes a simple expression given by 
\begin{eqnarray}
&&\hskip -9mm U(0)/J =-1+(1+\tanh 2K)\Big\{1-{\frac{1}{2}}\tanh R  \notag \\
&& -\Big[1+2(\cosh ^{3}2R-3\cosh 2R-2)I_{1}\Big]\coth 2R\Big\}
\label{U-ga-1}
\end{eqnarray}%
with $e^{-2R}=e^{2R_{1}}=(e^{4K}+1)/2$.

\begin{figure}[htbp]
\center{
\includegraphics [angle=0,width=8cm]{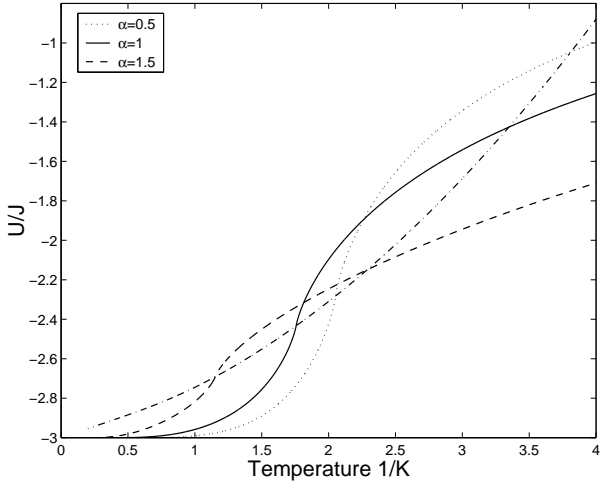}}
\caption{Energy $U$ versus temperature $T$ at fixed magnetic field 
$\protect\alpha$ for $\protect\gamma=-1$. The broken line indicates 
the critical energy $U_c$.}
\label{fig4}
\end{figure}

\begin{figure}[htbp]
\center{
\includegraphics [angle=0,width=8cm]{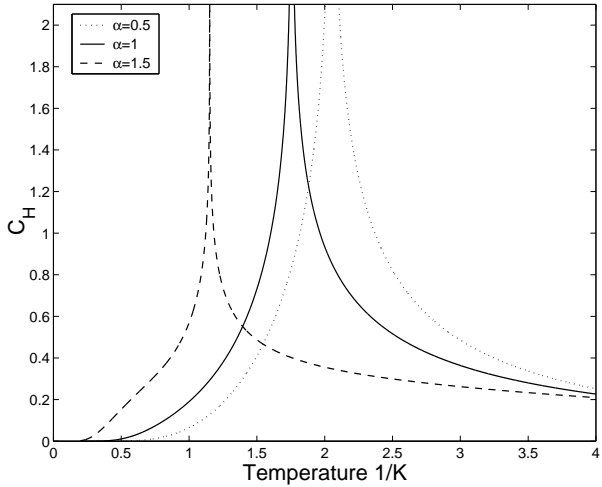}}
\caption{Specific heat $C_{H}$ as a function of temperature $T$ at fixed
magnetic field $\protect\alpha $ for $\protect\gamma =-1$.}
\label{fig5}
\end{figure}

\section{Derivation of $U_c(H)$}

Define an auxiliary variable 
\begin{eqnarray}
\varphi _{\alpha }(k)=\sin ^{-1}\sqrt{\frac{1-\alpha }{1-k^{2}}}\ >0\nonumber
\end{eqnarray}%
which gives, for $k^{2},r,s$ given in (\ref{krs}), 
\begin{eqnarray}
\varphi _{r}(k) &=&\sin ^{-1}\sqrt{1-\frac{x^{2}}{z^{2}}}  \notag \\
\varphi _{s}(k) &=&\sin ^{-1}\sqrt{\frac{(1-\omega ^{2})(z^{2}-x^{2})}{%
(z+\omega y)^{2}}}  \notag \\
\varphi _{s}(1) &=&\sin ^{-1}\sqrt{\frac{1-\omega }{1+\omega }\Big(1-\frac{%
x^{2}}{z^{2}}\Big)},\quad T=T_{c}(H)
\end{eqnarray}%
where the last line holds at the critical temperature $k=1$ or $y=z$.

For $k^{2}<r<1$ we have \cite{Abramowitz} 
\begin{eqnarray}
\Pi (r,k)=K(k)+{\frac{\pi }{2}}\sqrt{{\frac{r}{(1-r)(r-k^{2})}}}\ \Big[%
1-\Lambda _{0}(\varphi _{r},k)\Big] \notag
\end{eqnarray}%
where $\Lambda _{0}$ is Heuman's Lambda function 
\begin{equation}
\Lambda _{0}(\varphi ,k)={\frac{2}{\pi }}\Big[K(k)E(\varphi ,k^{\prime
})-[K(k)-E(k)]F(\varphi ,k^{\prime })\Big]
\end{equation}%
with $k^{\prime }=\sqrt{1-k^{2}}$ and 
\begin{eqnarray}
E(\varphi ,k) &=&\int_{0}^{\varphi }(1-k^{2}\sin ^{2}\theta )^{1/2}d\theta 
\notag \\
F(\varphi ,k) &=&\int_{0}^{\varphi }(1-k^{2}\sin ^{2}\theta )^{-1/2}d\theta .
\end{eqnarray}%
Particularly, 
\begin{equation}
\Lambda _{0}(\varphi ,1)={\frac{2}{\pi }}\varphi .
\end{equation}%
At the critical point $k\rightarrow 1$ or $y\rightarrow z$, we have 
\begin{eqnarray}
&&\hskip-9mm(z^{2}-y^{2})\Pi (r,k)=(z^{2}-y^{2})K(k)  \notag \\
&&\hskip4mm+{\frac{\pi z^{2}(z^{2}-x^{2})^{1/2}}{2x}}\Big(1-{\frac{2}{\pi }}%
\varphi _{r}\Big), \\
&&\hskip-9mm(z^{2}-y^{2})\Pi (s,k)=(z^{2}-y^{2})K(k)  \notag \\
&&\hskip4mm+{\frac{\pi (1+\omega )z^{2}(z^{2}-x^{2})^{1/2}}{2\sqrt{(1-\omega
)[(1-\omega )x^{2}-2\omega z^{2}]}}}\Big(1-{\frac{2}{\pi }}\varphi _{s}\Big),
\end{eqnarray}%
and 
\begin{eqnarray}
I_{2} &=&I_{1}-{\frac{1}{4x}}\Big(1-{\frac{2}{\pi }}\varphi _{r}\Big), 
\notag \\
I_{3} &=&{\frac{1}{1+\omega }}\Big[I_{1}  \label{Ii-critical} \\
&&+{\frac{\omega }{4\sqrt{(1-\omega )[(1-\omega )x^{2}+2\omega z^{2}]}}}\Big(%
1-{\frac{2}{\pi }}\varphi _{s}\Big)\Big].  \notag
\end{eqnarray}

Let 
\begin{eqnarray}
q=\sinh ^{2}R_{1}\overset{>}{<}{\frac{1}{2}},\quad |R_{1}|\overset{>}{<}|R|.\notag
\end{eqnarray}%
Then, at the critical temperature $T_{c}(H)$ 
\begin{eqnarray}
\omega &=&\frac{4q}{1+4q^{2}},\quad x=2(1+q),  \notag \\
y^{2} &=&z^{2}=4(1+q)^{2}(1+2q)/q,  \notag \\
\varphi _{r} &=&\sin ^{-1}\sqrt{\frac{1+q}{1+2q}}  \notag \\
\varphi _{s} &=&\sin ^{-1}\Bigg[\frac{|2q-1|}{2q+1}\sqrt{\frac{1+q}{1+2q}}\ %
\Bigg],  \notag \\
I_{2} &=&I_{1}-{\frac{1}{8(1+q)}}\Big(1-{\frac{2}{\pi }}\varphi _{r}\Big), 
\notag \\
I_{3} &=&{\frac{1+4q^{2}}{(1+2q)^{2}}}\Big[I_{1}  \notag \\
&&+{\frac{q}{2(1+q)|1-2q|(2q+3)}}\Big(1-{\frac{2}{\pi }}\varphi _{s}\Big)%
\Big],
\end{eqnarray}%
and 
\begin{eqnarray}
C_{1} &=&-{\frac{(q+1)(4q^{2}+8q-3)}{2q}}  \notag \\
&&\times \left( \coth 2R_{1}{\frac{\partial R_{1}}{\partial K}}-\coth 2R{%
\frac{\partial R}{\partial K}}\right) ,  \notag \\
C_{2} &=&-2(1+q)\left( \coth 2R_{1}{\frac{\partial R_{1}}{\partial K}}-\coth
2R{\frac{\partial R}{\partial K}}\right) ,  \notag \\
C_{3} &=&{\frac{(q+1)(2q-1)(2q+3)(2q+1)^{2}}{2q(4q^{2}+1)}}  \notag \\
&&\times \left( \coth 2R_{1}{\frac{\partial R_{1}}{\partial K}}-\coth 2R{%
\frac{\partial R}{\partial K}}\right) ,
\end{eqnarray}%
where 
\begin{eqnarray}
{\frac{\partial R_{1}}{\partial K}} &=&{\frac{2(e^{-4\gamma K_{c}}\sinh
4K_{c}-\gamma )}{e^{-4\gamma K_{c}}\cosh 4K_{c}-1}}+{\frac{2\gamma }{%
e^{-4\gamma K_{c}}-1}},  \notag \\
{\frac{\partial R}{\partial K}} &=&-{\frac{e^{-4\gamma K_{c}}(\sinh
4K_{c}-2\gamma \cosh ^{2}2K_{c})+\alpha \sinh 4\alpha K_{c}}{2(\sinh
^{2}2\alpha K_{c}+e^{-4\gamma K_{c}}\cosh ^{2}2K_{c})}}  \notag \\
&&+\alpha \tanh 2\alpha K_{c}
\end{eqnarray}%
where $K_{c}=J/kT_{c}(H)$. Combining these results, we obtain after some
algebra the following expression for the energy at the critical temperature: 
\begin{eqnarray}
&&\hskip-0.2cmU_{c}(H)/J=-\gamma -2\alpha \tanh 2\alpha K_{c}-{\frac{1}{%
2\sinh 2R_{1}}}{\frac{\partial R_{1}}{\partial K}}  \notag \\
&&\hskip1cm+{\frac{1}{2}}(2-\coth 2R){\frac{\partial R}{\partial K}}  \notag
\\
&&\hskip1cm+{\frac{1}{2\pi }}(\varphi _{r}\pm \varphi _{s})\Big(\coth 2R_{1}{%
\frac{\partial R_{1}}{\partial K}}-\coth 2R{\frac{\partial R}{\partial K}}%
\Big),  \notag \\
&&\hskip4cmq\overset{<}{>}{\frac{1}{2}}.  \label{UcH}
\end{eqnarray}

In the limit of $\gamma \rightarrow 0$ for which $R_{1}\rightarrow \infty $, 
$\varphi _{r}=\varphi _{s}=\pi /4$, we obtain 
\begin{eqnarray}
&&\hskip -0.5cm U_{c}(H)/J\Big|_{\gamma =0} =-(2-\sqrt{2})\alpha \tanh
2\alpha K_{c}  \notag \\
&&\hskip 1.5cm -\frac{1}{2-\sqrt{2}}{\frac{\sinh 4K_{c}+\alpha \sinh 4\alpha
K_{c}}{\sinh ^{2}2\alpha K_{c}+\cosh ^{2}2K_{c}}}  \label{Ucgamma0}
\end{eqnarray}%
which is the Fisher result \cite{Fisher60}. Particularly, for $H=0$, (\ref%
{Ucgamma0}) gives the value $-2\sqrt{1+\sqrt{2}}$.

\begin{figure}[tbp]
\center{
\includegraphics [angle=0,width=8cm]{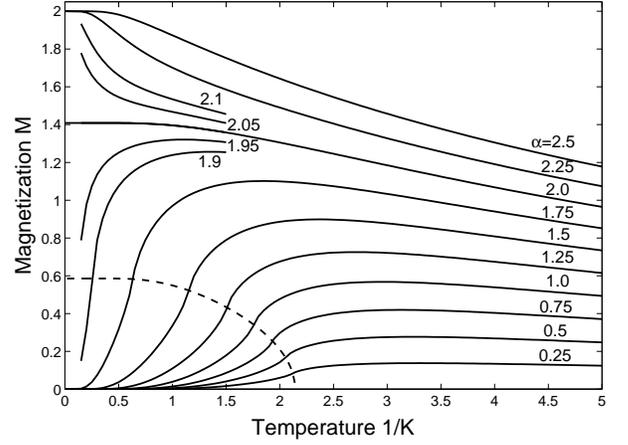}}
\caption{Magnetization $M$ versus temperature $T$ at fixed magnetic field
$\protect\alpha $ for $\protect\gamma =-1$. The broken line indicates 
$M_{c}(H)$.}
\label{fig6}
\end{figure}

\section{Magnetic properties}

For magnetic properties we take the derivative in (\ref{derivative}) with
respect to $H$ and obtain 
\begin{eqnarray}
{M(H)} &=&M_{c}(H)+\mathrm{const}\>|T-T_{c}(H)|\ln |T-T_{c}(H)|,  \notag \\
&&\hskip2cmT\rightarrow T_{c}(H)  \label{magH}
\end{eqnarray}%
where $M_{c}(H)$ is the magnetization at the critical temperature $T_{c}(H)$
given by 
\begin{eqnarray}
&& M_{c}(H)={\frac{1}{2}}\Big[1-{\frac{1}{\pi }}(\varphi _{r}\pm \varphi
_{s})\Big](1+e^{4R})\tanh 2\alpha K_{c},  \notag \\
&&\hskip 4cm q\overset{<}{>}{\frac{1}{2}}  \label{M-critical}
\end{eqnarray}

\begin{figure}[htbp]
\center{
\includegraphics [angle=0,width=8cm]{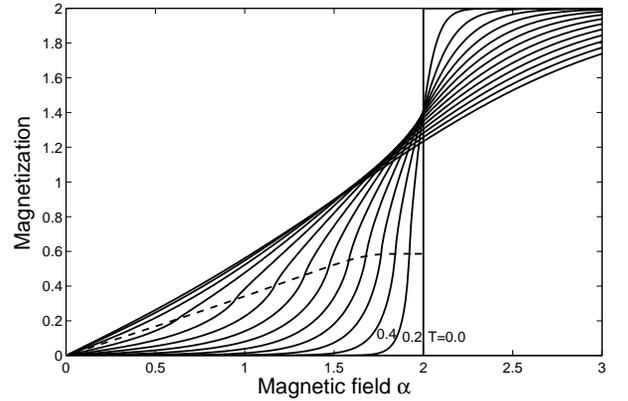}}
\caption{Magnetization $M$ versus magnetic field $\protect\alpha $ at fixed
temperature $T$ for $\protect\gamma =-1$. The broken line indicates 
$M_{c}(H)$.}
\label{fig7}
\end{figure}

The magnetization $M$ is plotted in Figs. 6 and 7 versus $T$ and $H$,
respectively, for $\gamma=-1$. To analyze $M(H)\equiv M(\gamma ,\alpha ,1/K)$
at low temperatures, we note that 
\begin{eqnarray}
M(\gamma ,\alpha ,0) &=&0,\quad \quad \alpha <1-\gamma  \notag \\
&=&2,\quad \quad \alpha >1-\gamma \ .
\end{eqnarray}%
The value of $M(\gamma ,\alpha ,0)$ at $\alpha =1-\gamma $ depends on how
the zero temperature $T=0$ is approached. To see what happens we write 
\begin{equation}
\alpha (K,\xi )\equiv 1-\gamma -{\frac{\xi }{4K}}\ln 2(\sqrt{2}+1)
\label{alp-path}
\end{equation}%
where $\xi $ is a parameter controlling the approach to $T=0$. Particularly,
the critical curve (\ref{K_c-big}) is indicated by $\xi =1$. Along (\ref%
{alp-path}) and $K\rightarrow \infty $, one has $R_{1}\simeq 2K\rightarrow
\infty $, $\tanh L\rightarrow 1$, and 
\begin{eqnarray}
e^{-4R}=1+[2(\sqrt{2}+1)]^{\xi },\quad K\rightarrow \infty .\nonumber
\end{eqnarray}%
After some algebraic manipulation, we find 
\begin{eqnarray}
m(\xi ) &\equiv &\lim_{K\rightarrow \infty }M\Big(\gamma ,\alpha (K,\xi ),1/K%
\Big)  \notag \\
&=&(e^{4R}+1)\Big[{\frac{1}{2}}+{\frac{\epsilon }{\pi }}(1-k)K(k)\Big]
\end{eqnarray}%
with $k=\sinh ^{2\epsilon }2R$. Here $\epsilon =1$ for $\xi \leq 1$ while $%
\epsilon =-1$ for $\xi >1$. We have 
\begin{eqnarray}
m(\xi ) &=&0\hskip4.4cm\xi =\infty  \notag \\
&=&2-\sqrt{2}\hskip3.5cm\xi =1  \notag \\
&=&\frac{3}{4}+\frac{21}{16\pi }{K\Big(\frac{1}{8}\Big)} \simeq 1.408\ 836
\quad \xi =0  \notag \\
&=&2\hskip4.4cm\xi =-\infty .
\end{eqnarray}

\begin{figure}[htbp]
\center{
\includegraphics [angle=0,width=8cm]{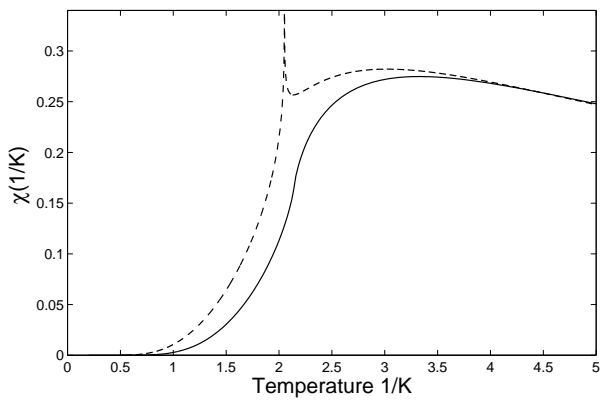}}
\caption{Susceptibility as a function of temperature for $\protect%
\gamma =-1$. Solid line $H=0$. Broken line $H=J\ (\protect\alpha =1/2$).}
\label{fig8}
\end{figure}

The susceptibility $\chi (\gamma ,\alpha ,K^{-1})$, which is the further
derivative of (\ref{magH}) with respect to $H$, diverges logarithmically at $%
T_{c}(H)$ for $H\neq 0$. The susceptibility $\chi $ is plotted in Fig. 8
versus temperature $T$ for $\gamma =-1$ and $\alpha =0,1/2$. The zero-field
susceptibility $\chi (\gamma ,0,K^{-1})$ is continuous. For example one has
for $\gamma =-1$, 
\begin{eqnarray}
&&\hskip-5mm\chi (-1,0,K^{-1})=2K-{\frac{K}{2}}(1-e^{4R})\Bigg\{2-{\frac{3}{2%
}}\coth 2R  \notag \\
&&\hskip1.8cm+\coth 2R\Big[2\cosh 2R(3-\sinh ^{2}2R)+6\Big]I_{1}  \notag \\
&&\hskip1.8cm+\sinh 4R\>I_{2}\Bigg\}  \label{chi0-ga-1}
\end{eqnarray}%
where $e^{-2R}=(e^{4K}+1)/2$. The expression (\ref{chi0-ga-1}) bears no
direct relation with the critical energy (\ref{U-ga-1}) as found to exist in
the case of $\gamma =0$ \cite{Fisher60}. By direct differentiation of (\ref%
{M-critical}) and making use of (\ref{TcH}), we obtain the critical
susceptibility 
\begin{eqnarray}
\chi (\gamma ,0,K_{c}^{-1}) &=&{\frac{1}{2}}K_{c}\Big[1-{\frac{1}{\pi }}%
(\varphi _{r}\pm \varphi _{s})\Big]  \notag \\
&&\times (1+e^{4\gamma K_{c}}\cosh ^{-2}2K_{c}),\quad q\overset{<}{>}{\frac{1%
}{2}}.
\end{eqnarray}

\section*{Acknowledgment}

We thank R. Shrock for his interest and a critical reading of the
manuscript. Work has been supported in part by NSF Grants PHY-0098801 (WTL)
and DMR-9980440 (FYW).


\begin{thebibliography}{9}
\bibitem{corrections} Expressions of $I_1,I_2,I_3$ given just above Eq. (41)
in \cite{Fan70} contain the following typos: The arguments $g$ of the
elliptical integrals should read $q$. In the expression of $I_1$ replace $%
(z^2-x^2) \to (z^2-x^2)^{1/2}$, in the expression of $I_2$ multiply the
whole expression by $1/z$, and in the expression of $I_3$ replace the first $%
g$ by $y$.
\end{thebibliography}
\end{document}